\documentclass{tMPH2e}

\usepackage[colorlinks = true, linkcolor = black, citecolor = black, urlcolor = blue]{hyperref}
\usepackage{breakurl}

\begin{document}
\doi{10.1080/0026897YYxxxxxxxx}
 \issn{1362–3028}
\issnp{0026–8976}
\jvol{00}
\jnum{00} \jyear{2009} 

\markboth{R. Matthews, A.A. Louis and J.M. Yeomans}{Molecular Physics}

\articletype{RESEARCH ARTICLE}

\title{Confinement of knotted polymers in a slit}

\author{R. Matthews$^{\ast}$\thanks{$^\ast$Corresponding author. Email: r.matthews1@physics.ox.ac.uk
\vspace{6pt}}, A.A. Louis and J.M. Yeomans\\\vspace{6pt}  {\em{Rudolf Peierls Centre for Theoretical Physics, 1 Keble Road, Oxford 0X1 3NP, England}}
\\\vspace{6pt}\received{v4.5 released September 2009} }

\maketitle

\begin{abstract}
We investigate the effect of knot type on the properties of a ring polymer confined to a slit. For relatively wide slits, the more complex the knot, the more the force exerted by the polymer on the walls is decreased compared to an unknotted polymer of the same length. For more narrow slits the opposite is true. The crossover between these two regimes is, to first order, at smaller slit width for more complex knots.  However, knot topology can affect these trends in subtle ways. Besides the force exerted by the polymers, we also study other quantities such as the monomer-density distribution across the slit and the anisotropic radius of gyration. \bigskip

\bigskip

\end{abstract}

\section{Introduction}

Whilst knots are generic, and any polymer of sufficient length in solution is very likely to be knotted~\cite{sumners}, they have proven to have particular relevance in biological systems. The DNA contained in the cells of living organisms is typically both long and confined: the genome of $E.$ $coli$, a well studied bacterium~\cite{berg_biochem}, is about $3\times10^4$ persistence lengths~\cite{berg_biochem} ($\sim 1.5 mm$) and is contained in a cell with a typical volume $\sim \mu m^3$~\cite{kubitscheck}. Although the DNA is organized using supercoiling and proteins (particularly in eukaryotes)~\cite{woldringh}, confinement makes entanglements likely~\cite{tesi,michels}, particularly during processes such as replication~\cite{sogo}. Knots in DNA can prevent transcription~\cite{portugal}, a process necessary for the production of proteins. Indeed, there is a family of enzymes, the topoisomerases~\cite{watt}, one of whose functions is to control knotting. Evidence of highly confined DNA knotting has also been found in bacteriophages~\cite{arsuaga}, viruses that infect bacteria. These knots may affect the ejection speed of a bacteriophage's DNA~\cite{matthews,marenduzzo}. Given its relevance in biology, it is important to understand the interplay of knotting and confinement.

The confinement of linear polymers in a slit, between two infinite, parallel walls has been well studied. The simplest model is to consider a single chain with walls that provide only geometric constraints. For $D > R$, where $D$ and $R$ are slit and polymer size respectively, the chain will maintain a shape similar to that seen without confinement. For $D < R$, however, significant deformation occurs. A scaling form for the free energy of such a chain for  $D < R$ was given by Daoud and de Gennes~\cite{daoud} using a blob picture, in which the polymer is taken to form a series of independents blobs of size determined by $D$, within each of which the polymer behaves as if it is free. They predicted
\begin{equation}
\frac{F_{conf}}{k_B T} \approxeq N \left(\frac{a}{D}\right)^{1/\nu},
\label{eq:free_energy_confine_de_gennes}
\end{equation}
where $N$ is the number of monomers in the chain, $a$ is monomer size and $\nu$ is a three dimensional exponent $\approx 3/5$. Predictions made by Daoud and de Gennes for the polymer size were verified by Webman, Lebowitz and Kalos using simulations~\cite{webman}. Later, it was predicted that the ratio between the monomer density by the wall and the force exerted on the wall is universal, independent of microscopic details~\cite{eisenriegler}. Again, attempts have been made to verify this prediction by simulation, see ref.~\cite{hsu} and references therein. Whilst the majority of simulation work has used Monte Carlo methods, a recent study has also compared the results of Molecular Dynamics simulations to scaling predictions~\cite{dimitrov}. Other work has considered issues such as attractive walls~\cite{brak}.

By comparison to the case of linear polymers, there has been little work on knotted polymers in slits and basic issues remain open. Tesi $et$ $al.$~\cite{tesi} looked at the probability that a knot will be found in a slit-confined polymer. They found that for relatively wide slits the probability increased as the slit width decreased. However, for narrower slits, the probability peaked and eventually became less than the value for wide slits. Janse van Rensburg~\cite{rensburg} applied both analytical calculations and a Monte Carlo approach that samples configurations with the same topology but varying $N$. The change of the average value of $N$ with slit width was investigated. Knotted polymers were found to expand as the width increased, in contrast to unknotted polymers whose size showed a plateau after a certain width was reached. In ref.~\cite{farago}, the ends of knotted and unknotted polymers were attached to parallel plates, separated so as to stretch the chains. It was found that the forces exerted on knotted polymers deviated from scaling predictions. The results were used to deduce an estimate of how the knot size scales with polymer length.

Motivated by the common occurrence of confined knotted polymers in biology, we use a coarse-grained model to gain insight into the behaviour of knotted ring polymers in a slit, concentrating on the force exerted on the walls. We then go on to look at other quantities, such as the radius of gyration and the monomer density distribution, that help us to interpret the trends seen.

We find that the effect of knotting differs depending on the degree of confinement of the polymer. We find that for narrow slits, more complex knots exert higher forces on the walls, whereas for relatively wide slits, the opposite is true. We relate the forces to the monomer densities near the walls and interpret the results in terms of the effect of knots on the ability of the polymer to spread out in the slit.  

\section{Model}
\label{sec:knot_slit_model}

We use a standard, bead-spring, polymer model~\cite{grest}. Excluded volume is included through a truncated Lennard-Jones potential:
\begin{equation}
\begin{array}{ll}
V_{EV} (x) = 4\epsilon\left[\left(\frac{\sigma}{x}\right)^{12}-\left(\frac{\sigma}{x}\right)^{6}+\frac{1}{4}\right], & \;\;\;\; x \leq 2^{1/6}\sigma , \\
V_{EV} (x) = 0, & \;\;\;\; x > 2^{1/6}\sigma , \\
\end{array}
\label{eq:LJ_trunc}
\end{equation}
where $\sigma$ is the size of a bead, $\epsilon$ is an energy scale and $x$ is the bead separation. Neighbouring beads are connected with FENE springs~\cite{warner}
\begin{equation}
V_{FENE} = -\frac{k X_0^2}{2}\ln\left[1-\left(\frac{x}{X_0}\right)^2\right],
\label{eq:FENE}
\end{equation}
where $X_0$, the maximum extension, is set to $1.5\sigma$ and $k$ is chosen as $30 \epsilon / \sigma^2$.

The polymer motion is simulated using the Langevin equation~\cite{lemons}:
\begin{equation}
m \mathbf{\ddot{x}}_i = \mathbf{f}_i - \zeta \mathbf{\dot{x}}_i + \sqrt{2 \zeta k_B T} 
 \mathbf{r}_i(t),
\label{eq:lang_eqn}
\end{equation}
where $\mathbf{x}_i$ is the position of the $i^{th}$ bead, $\mathbf{f}_i$ is the total force acting on it, $m$ is the bead mass, $\zeta$ is a friction constant and $k_B T$ is the temperature, which we set to $\epsilon$. $\mathbf{r}_i(t)$ are random vectors that satisfy
\begin{eqnarray}
\nonumber  \left< \mathbf{r}_i (t) \right>  &=& 0,
\\  \left<\mathbf{r}_i(t)\mathbf{r}_j(t') \right> &=& \delta_{ij}\delta(t-t')\mathcal{I},
\label{noise_correl}
\end{eqnarray}
where $\mathcal{I}$ is the identity matrix. The second and third terms in Eq.~(\ref{eq:lang_eqn}) represent the drag due to the solvent and Brownian noise respectively. The parameters may be combined to give a convenient unit of simulation time, $t_0 = \sqrt{\left(\sigma^2 m / \epsilon\right)}$. We integrate the motion using a velocity Verlet algorithm~\cite{frenkel} with a timestep of $\Delta t  = 0.01$.

The polymer is placed in a slit between two parallel plane walls, modelled by the same excluded volume potential as for the bead-bead interactions, Eq.~(\ref{eq:LJ_trunc}). The surfaces of the walls are taken as lying at the point at which the potential is truncated. We consider slit widths from $D = 4\sigma$ to $D = 18 \sigma$. This range was chosen, through initial exploratory simulations, to show the crossover between the regimes where more complex knots are harder or easier to confine. The simulations resemble those of Dimitrov $et$ $al.$~\cite{dimitrov}, who considered confined, linear polymers.

We choose a polymer length, $N = 300$, which is sufficiently short that the knots will significantly affect the polymer properties. It should be borne in mind that, since knots in polymers in a good solvent are weakly localized~\cite{orlandini2}, different results would be obtained for other values of $N$. The results do, however, give a qualitative picture of how a knot, $which$ $is$ $of$ $significant$ $length$ $compared$ $to$ $N$, will affect a polymer in a slit.

\begin{figure}
\begin{center}
\includegraphics[scale=0.25]{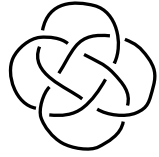}
\caption{\label{fig:8_18_diagram} Knot topology for $8_{18}$, based on a diagram from ref.~\cite{livingston}.}
\end{center}
\end{figure}

We consider linear and unknotted ring polymers (denoted $0_1$) and also polymers with knot types  
$3_1$, $6_1$ and $9_1$. $C_p$ is a standard notation~\cite{orlandini}, where $C$ gives the minimum number of crossings in any projection onto a plane, the essential crossings. $p$ is used to distinguish knots with equal $C$. $3_1$ and $9_1$ are torus knots; they both belong to a group of knots with similar topology. $6_1$ is an even-twist knot. In addition, we consider two other knots. Firstly, we include a twelve-crossing even-twist knot. We label it $12_1$, although the standard notation only extends up to $C = 10$. Secondly, we also look at $8_{18}$, which is chosen because in initial simulations it was found to have an unusually spherical average instantaneous shape~\cite{aronovitz}. In Fig.~\ref{fig:8_18_diagram}. we show a knot diagram~\cite{livingston} for $8_{18}$. The polymer model used should prevent chain crossings~\cite{grest} and we verified that the knot type remained constant by calculating the Alexander polynomial~\cite{orlandini}, $A_k(t)$, at $t = -1$ and $t = -2$.

The radius of gyration of the unconfined linear polymer, $R_G(\textrm{linear}) =  14.2\pm0.1\sigma$. For the ring polymers with different knots, $R_G(0_1) = 10.65\pm0.03\sigma$, $R_G(3_1) = 9.01\pm0.02\sigma$, $R_G(6_1) = 7.78\pm0.01\sigma$, $R_G(8_{18}) = 6.86\pm0.01\sigma$, $R_G(9_1) = 7.28\pm0.01\sigma$ and $R_G(12_1) = 6.90\pm0.01\sigma$. However, it should be kept in mind that the instantaneous configurations of linear polymers are not spherical, but rather one can define 3 axes that characterize a prolate shape~\cite{aronovitz}. We investigated the anisotropic instantaneous distribution of knotted polymer shapes, which tend to be more spherical than linear polymers~\cite{rawdon}, although here there is no universal behaviour because it depends on the size of the knot compared to the length of the polymer.

We note that for some of the more complex knots, an increase in the internal energy of the chain larger than the standard error was observed: for narrow slits and complex knots the probability of collision increases and so the amount of overlap between beads is higher, leading to an internal energy increase due to the excluded volume interaction. The largest such increase, $\Delta U = 0.56 \pm 0.06 k_B T$, was seen for $12_1$ between $D = 8\sigma$ and $D = 4\sigma$. In the middle of this range, the force on the confining walls was $7.734\pm0.003 k_B T / \sigma$. This suggests that the internal energy change accounts for $\lesssim 1\%$ of the free energy increase of the chains, with the rest being due to entropy. We are thus confident that this effect, which depends on the specific potential chosen, does not affect our conclusions, which are qualitative.

An initialisation period of $10^5 t_0$ was allowed before results were recorded for $10^7 t_0$. Relaxation was slower in the two free directions than the confined direction and also depended on $D$ and knot type~\cite{quake}: depending on the particular system, the timescale may vary with knot complexity in different ways~\cite{matthews2}. The relaxation time in the free directions, taken as the time for the auto-correlation function of the relevant component of the radius of gyration to decay to $1/e$, was typically on the order of $10^2 t_0$ for the knotted polymers and even for the slowest case ($D = 4 \sigma$, linear polymer) was clearly less than $10^4 t_0$.

Errors were estimated from the variance of the relevant quantities, initially assuming that all data points were independent. During the simulations, data was recorded every $100 t_0$. If the estimate of the relevant correlation time, $t_C$, was seen to be significantly greater than $100 t_0$ then the estimated error was increased by a factor of $\sqrt{2 t_C / 100 t_0}$~\cite{frenkel}. 

\section{Results}

\begin{figure}
\begin{center}
\includegraphics[scale=0.25]{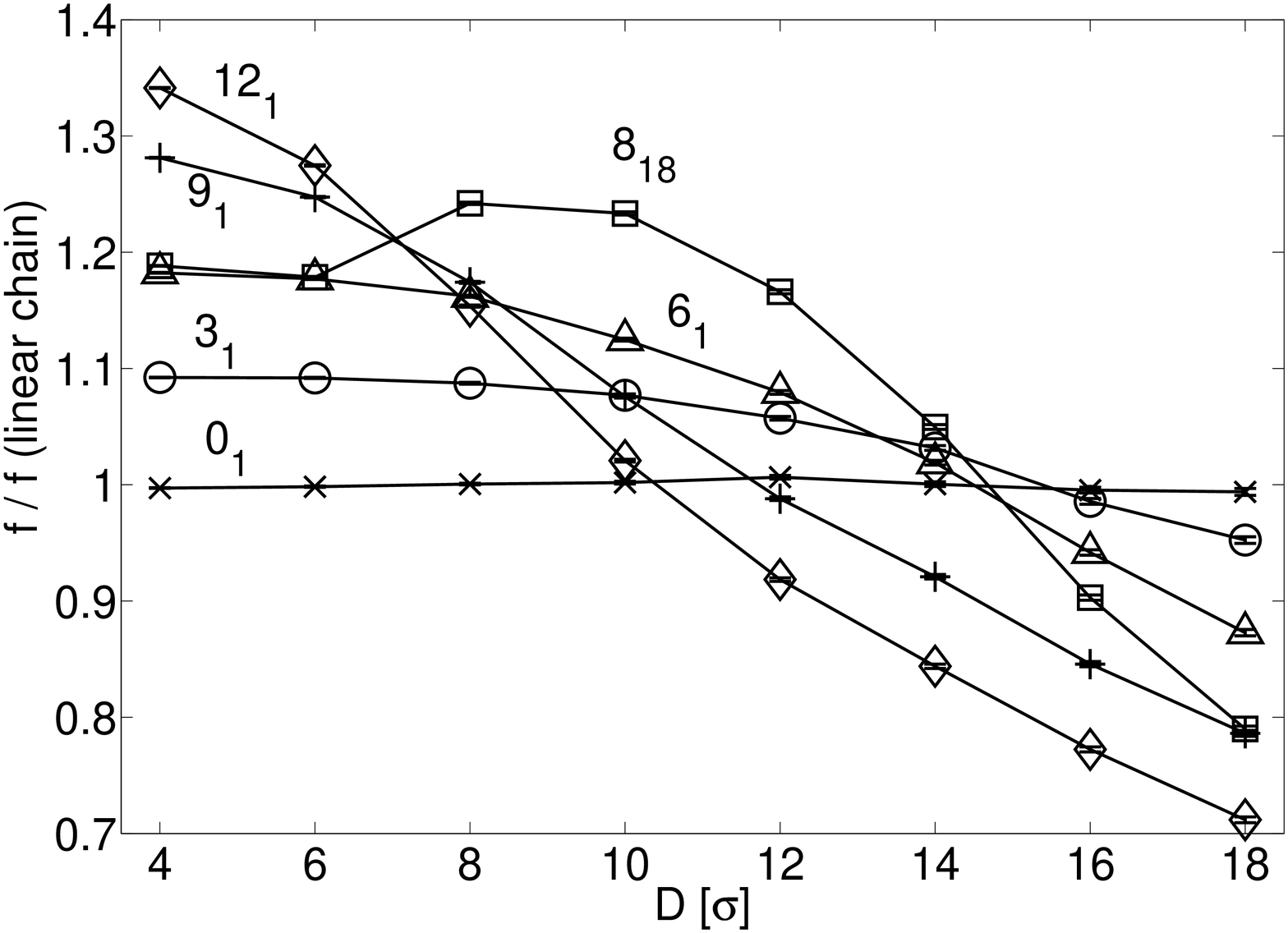}
\caption{\label{fig:force_knot_linear_ratio_width} The average force, $f$, exerted on the walls of a confining slit by polymers with different knot types -- $0_1$ ($\times$), $3_1$ ($\bigcirc$), $6_1$ ($\triangle$), $8_{18}$ ($\square$), $9_1$ (+) and $12_1$ ($\lozenge$) -- divided by the values for a linear chain, as a function of slit width, $D$. Errorbars are plotted but for most points they are much smaller than the symbols.}
\end{center}
\end{figure}

We study the average force, $f$, exerted on the confining walls by the polymer as function of $D$ for different knot types. The average force for knotted and unknotted rings is plotted as a ratio to the value for a linear chain in Fig.~\ref{fig:force_knot_linear_ratio_width}.  

We first consider the forces exerted by linear and unknotted ring polymers, which are very similar over the range of slit widths investigated. The scaling picture of a confined polymer is based on it forming a series of blobs of dimension $\sim D$~\cite{daoud}. For small widths, both linear and ring polymers should form the same number of blobs. However, since linear polymers have a larger radius of gyration, it would be expected that, for wide enough slits ($D \gg R_G$), the forces exerted by the unknotted ring would decrease below those for linear polymers. At $D = 30\sigma$, the component of the radius of gyration in the direction of confinement was $5.24\pm0.01\sigma$ for a linear polymer, compared to $5.14\pm0.01\sigma$ for an unknotted ring, and the force exerted by the unknotted polymer was $0.886 \pm 0.006$ of the linear polymer value. The difference reflects the fact that $R_G(0_1) < R_G(\textrm{linear})$.

The topological constraints of a knot that is relatively large compared to the polymer prevent the formation of independent blobs so it is expected that results for knotted rings will deviate from those for linear and unknotted polymers. Fig.~\ref{fig:force_knot_linear_ratio_width} shows that, for narrow slits, polymers with more complex knots exert a higher force, whilst for wide slits the opposite is true. The crossover occurs at smaller slit widths for more complex knots, likely because they are more compact~\cite{quake}. $8_{18}$, picked because it is unusually spherical, shows, uniquely amongst the knots studied, non-monotonic behaviour for the force ratio as a function of $D$.

\begin{figure}
\begin{center}
\includegraphics[scale=0.25]{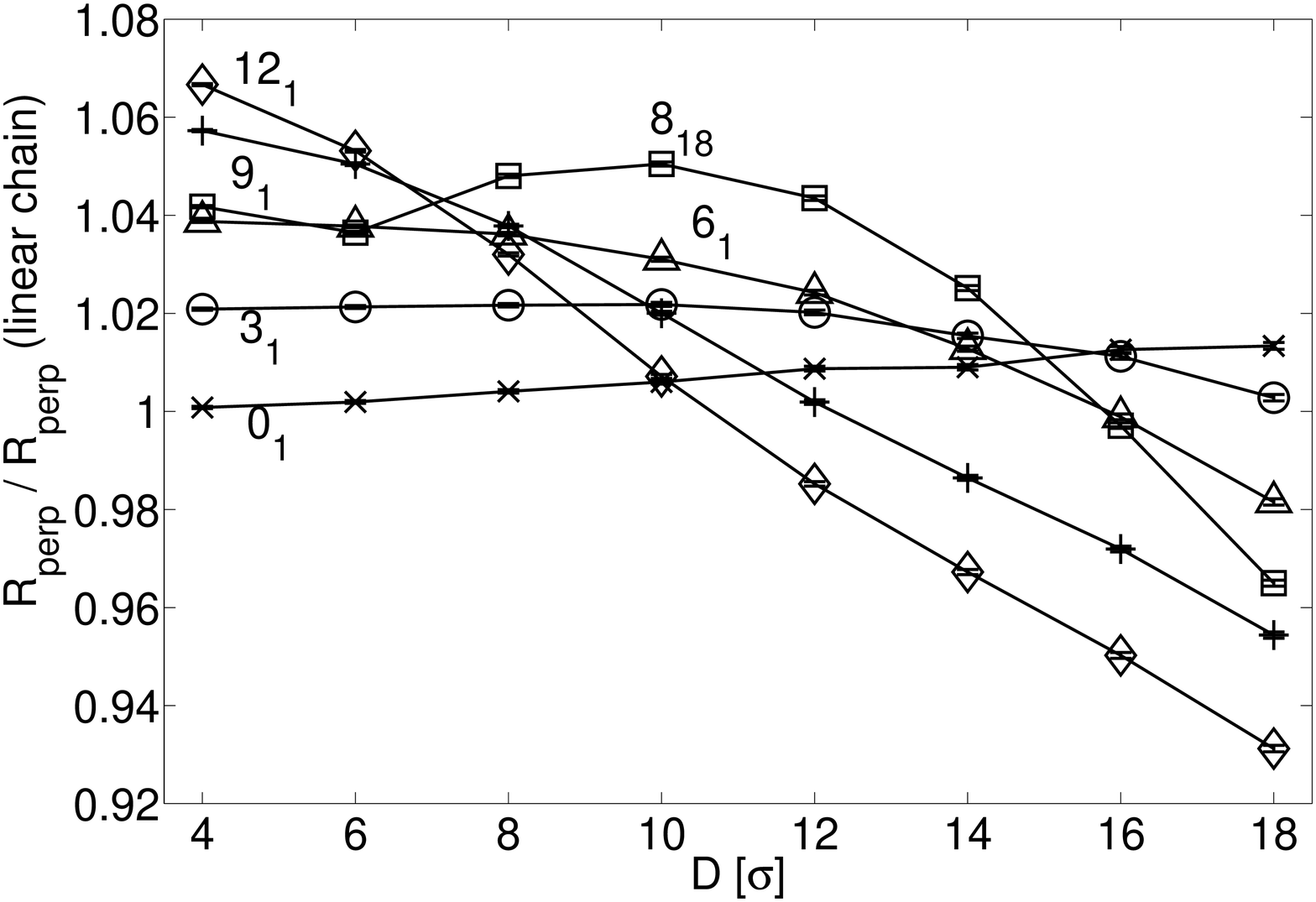}
\caption{\label{fig:R_perp_knot_linear_ratio_width} The radius of gyration, in the confined direction, $R_{perp}$, of polymers in a slit with different knot types -- $0_1$ ($\times$), $3_1$ ($\bigcirc$), $6_1$ ($\triangle$), $8_{18}$ ($\square$), $9_1$ (+) and $12_1$ ($\lozenge$) -- divided by the values for a linear chain, as a function of slit width, $D$.}
\end{center}
\end{figure}

We next compare the radius of gyration of the polymers, focussing on the component in the confined direction, $R_{perp}$. In Fig.~\ref{fig:R_perp_knot_linear_ratio_width}, results are plotted as ratios to the values for a linear polymer. The results for $R_{perp}$ show very similar trends to those seen for the forces: as may be appreciated by comparing Figs.~\ref{fig:force_knot_linear_ratio_width} and~\ref{fig:R_perp_knot_linear_ratio_width}, the ordering of the different knot types is almost identical for all $D$. 


\begin{figure*}
\begin{center}
\begin{minipage}{140mm}
\subfigure[]{
\resizebox*{6.5cm}{!}{\includegraphics{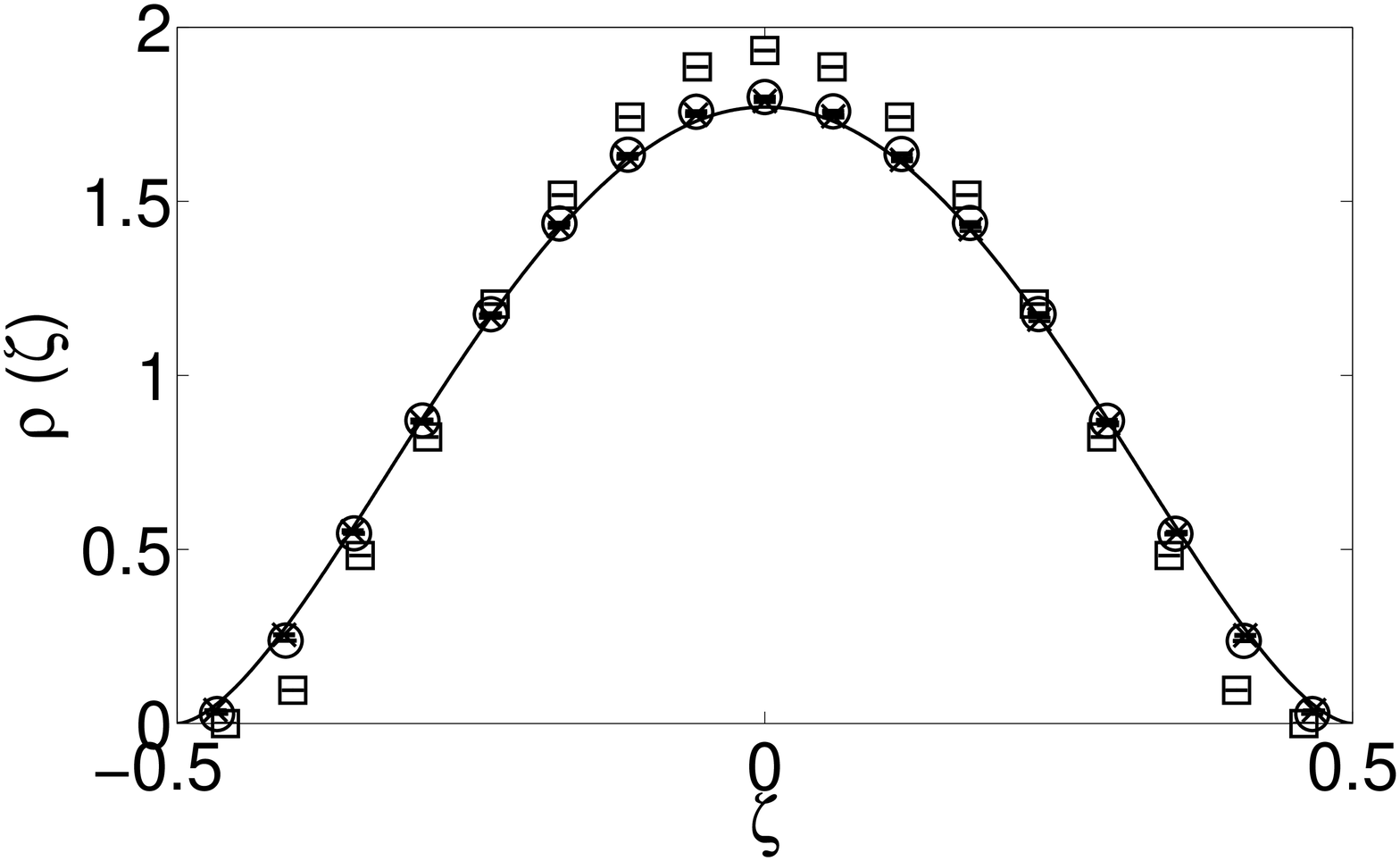}}}\hspace{5pt}
\subfigure[$D = 4\sigma$]{
\resizebox*{6.5cm}{!}{\includegraphics{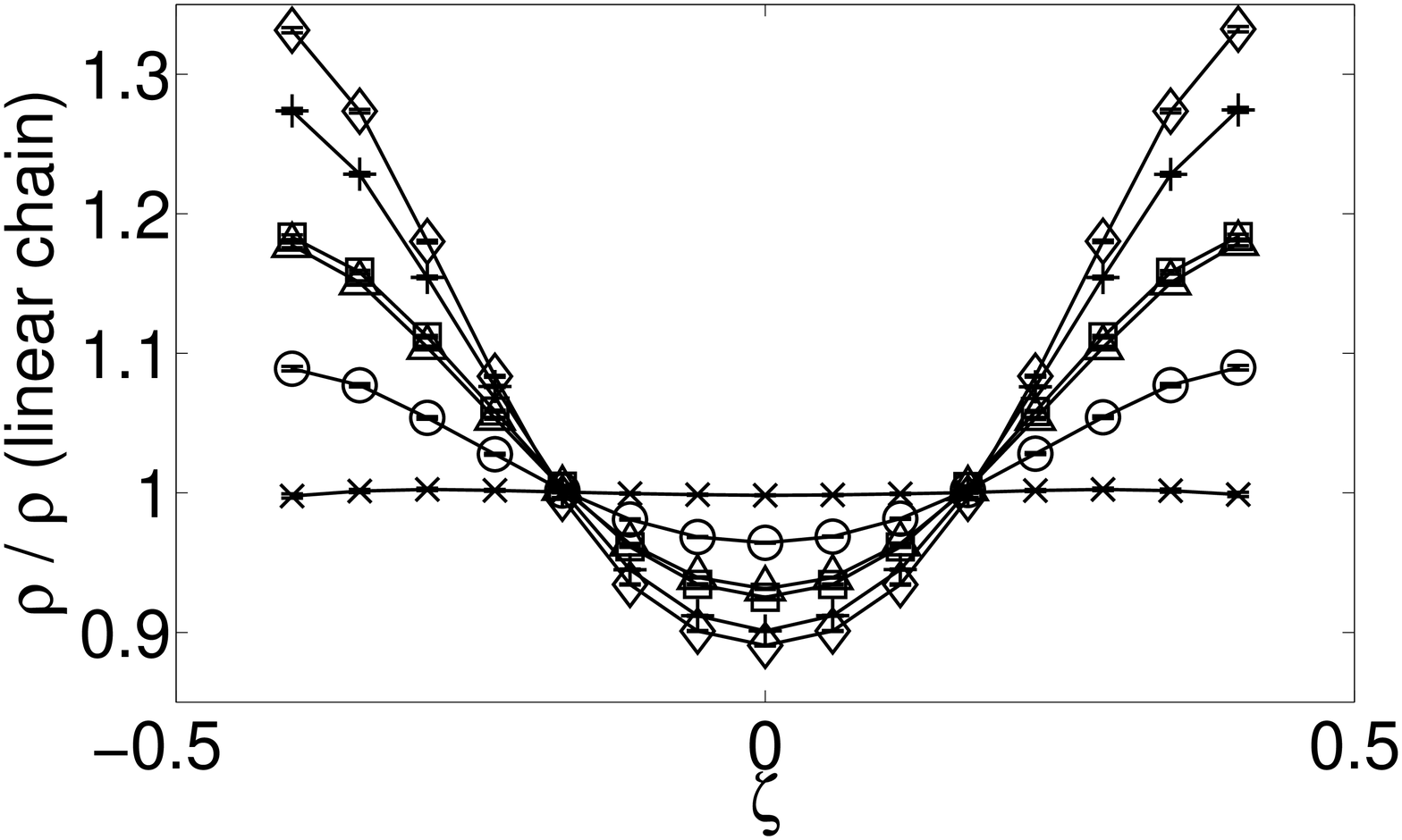}}}

\subfigure[$D = 12\sigma$]{
\resizebox*{6.5cm}{!}{\includegraphics{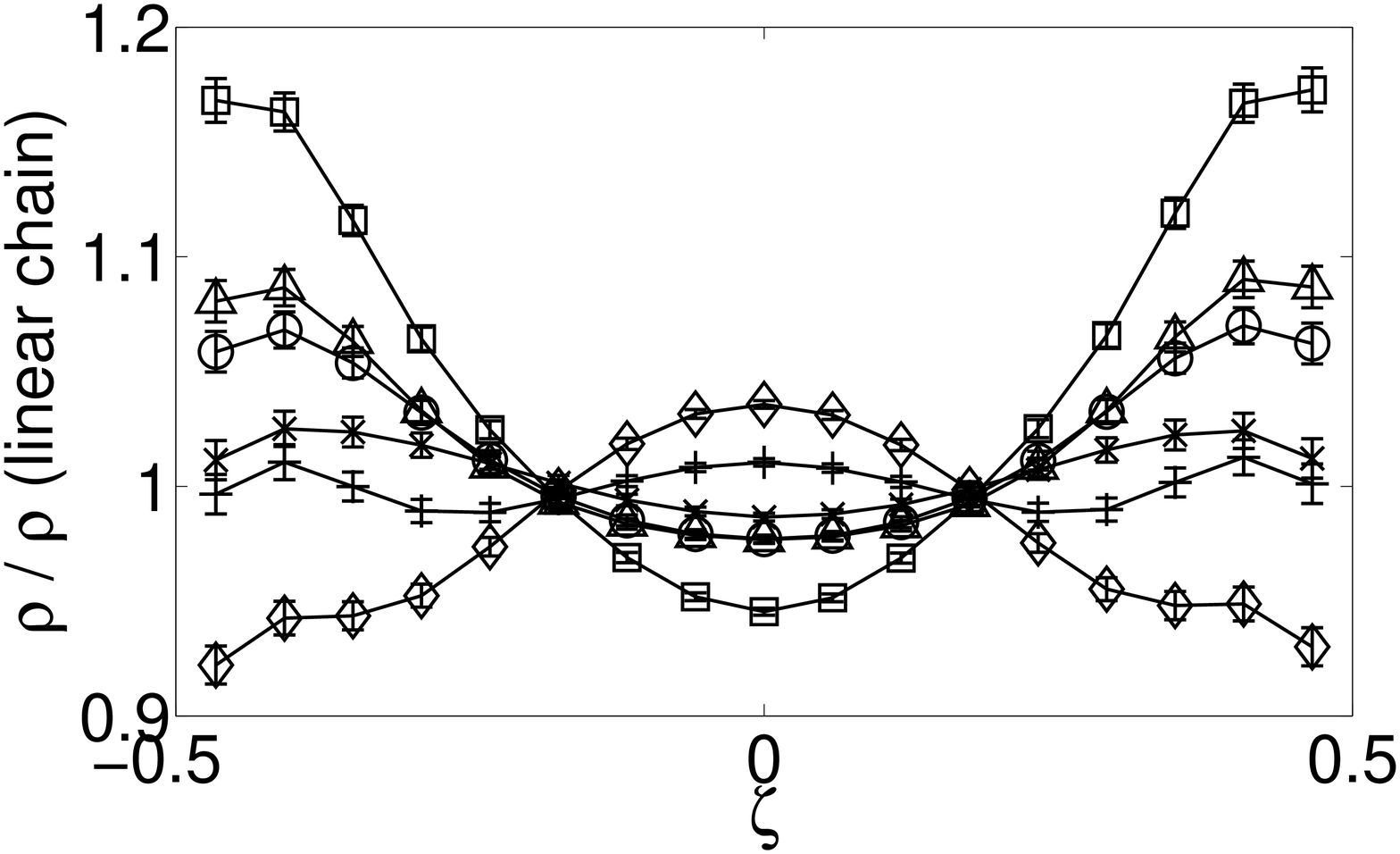}}}\hspace{5pt}
\subfigure[$D = 18\sigma$]{
\resizebox*{6.5cm}{!}{\includegraphics{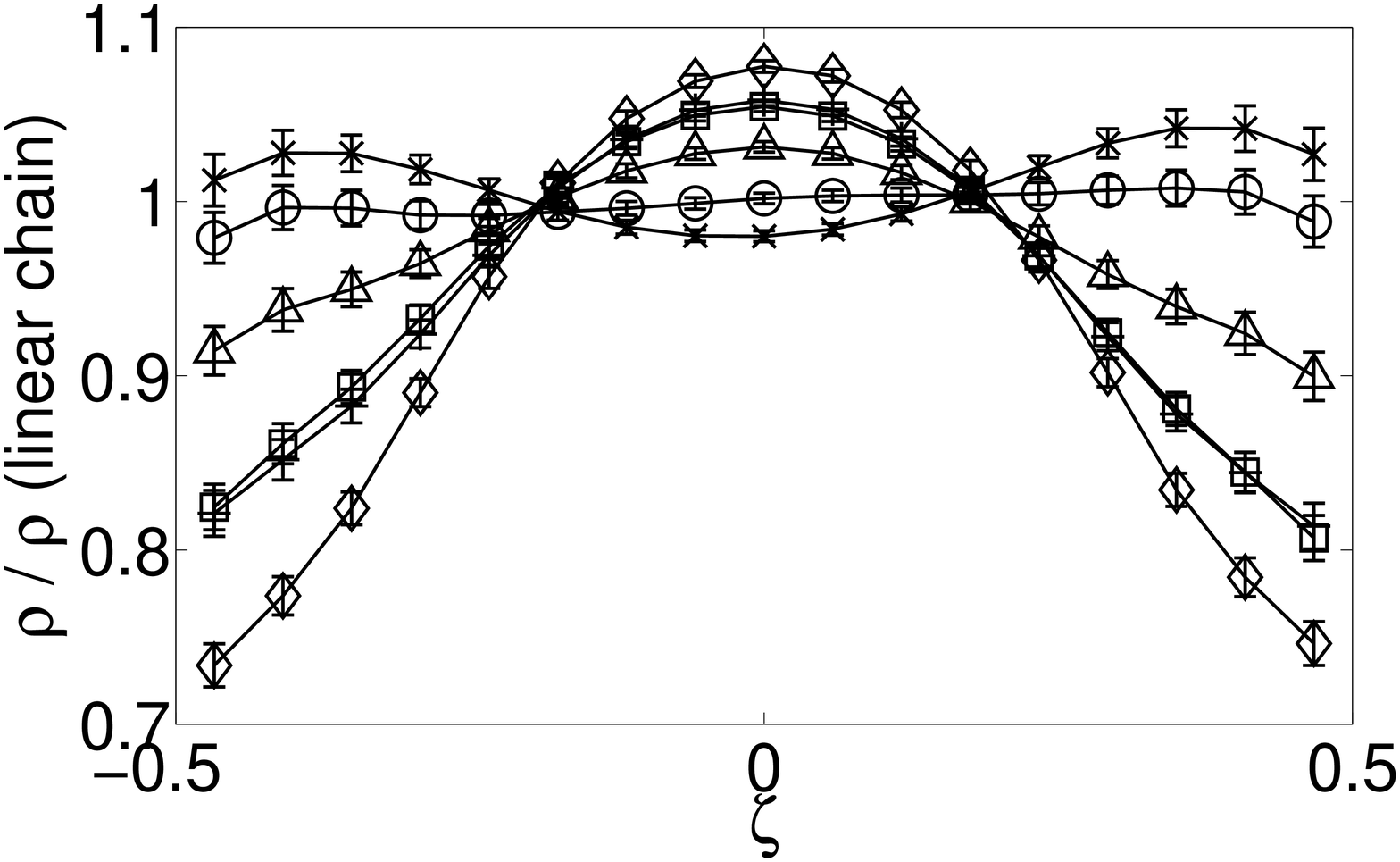}}}
\caption{\label{fig:mon_dens_multi} Monomer densities as a function of $\zeta = z / (D + \delta a)$, where $z$ is the distance from the centre of the slit and $\delta$ is a correction necessary to achieve agreement with scaling formulae when $D$ is comparable to $a$~\cite{hsu,dimitrov}. (a) Monomer density across the slit for $D = 4 \sigma$ ($\square$), $D = 12 \sigma$ ($\bigcirc$) and $D = 18\sigma$ ($\times$) for linear polymers. The solid line is a scaling formula~\cite{hsu,dimitrov}. We use a scaling exponent $\nu = 0.588$~\cite{dimitrov} and a correction, $\delta = 1.38$, found to give the best fit to the data. (b)-(d) Monomer density as function of the position across the slit for various knot types: $0_1$ ($\times$), $3_1$ ($\bigcirc$), $6_1$ ($\triangle$), $8_{18}$ ($\square$), $9_1$ (+) and $12_1$ ($\lozenge$). Results, for (b) $D = 4\sigma$, (c) $D = 12\sigma$ and (d) $D = 18\sigma$, are plotted as ratios to the values for linear polymers.}%
\end{minipage}
\end{center}
\end{figure*}

To help interpret the results for $f$ and $R_{perp}$, we examine the monomer density as a function of position across the slit for various $D$. The results are shown in Fig.~\ref{fig:mon_dens_multi}. Firstly, in Fig.~\ref{fig:mon_dens_multi}(a), we compare the results for linear polymers to the following scaling formula for monomer density~\cite{hsu,dimitrov}: 
\begin{equation}
\rho (\zeta) = \frac{\Gamma(2 + 2/\nu)}{ (\Gamma(1 + 1/\nu))^2 } (1/4 - \zeta^2)^{1/\nu},~
\end{equation}
where $\zeta$ is the position across the slit divided by $D + \delta a$. $\delta$ is a correction, necessary  when $D$ is comparable to $a$~\cite{hsu,dimitrov}. Assuming a scaling exponent, $\nu = 0.588$~\cite{dimitrov}, we find the best fit to the data using $\delta = 1.38$. The results show reasonable agreement with the scaling formula, although clear differences are seen, particularly for $D = 4\sigma$. Given that, for this width, $D$ is only a few times bigger than the bead size, it is unsurprising that such deviations are seen. We also checked our results by comparing the forces for linear polymers to a scaling formula derived from Eq.~(\ref{eq:free_energy_confine_de_gennes}). As in the work of Dimitrov {\it et al.}~\cite{dimitrov}, we found good agreement using the same value of $\delta$ as for the monomer density.

In Figs.~\ref{fig:mon_dens_multi}(b)-(d), we compare monomer densities across the slit for different knot types for $D = 4\sigma$, $12\sigma$ and $18\sigma$, by plotting them as ratios to the results for linear polymers. These simulations show that knots change how the polymer is distributed across the slit in a different way for different slit widths. For wide slits, $D \gtrsim 14\sigma$, the constraints of more complex knots make the polymer more compact~\cite{quake} leading to a lower monomer density at the walls. For narrower slits, $D \lesssim 8\sigma$, however, the more complex the knot the higher the wall density. The value of the monomer density is expected to control the force exerted by the polymer and comparison of Figs.~\ref{fig:force_knot_linear_ratio_width} and~\ref{fig:mon_dens_multi} show a close correlation between the two quantities.

Similarly, the trends for the radius of gyration across the slit shown in Fig.~\ref{fig:R_perp_knot_linear_ratio_width} can be explained in terms of the results in Fig.~\ref{fig:mon_dens_multi}: for narrower slits the monomers tend to lie closer to the walls for more complex knots leading to an increase in $R_{perp}$, whereas for wider slits the reverse is true. 

\begin{figure}
\begin{center}
\includegraphics[scale=0.25]{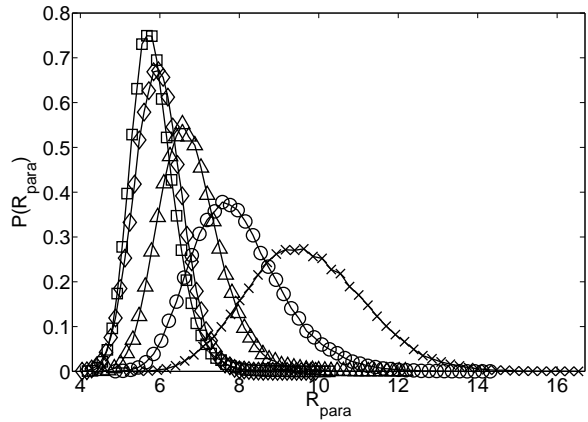}
\caption{\label{fig:R_para_dists} Distribution of the (instantaneous) radius of gyration in the two free directions, $R_{para}$, for ring polymers in a slit of $D = 18\sigma$ with different knots: $0_1$ ($\times$), $3_1$ ($\bigcirc$), $6_1$ ($\triangle$), $8_{18}$ ($\square$) and $12_1$ ($\lozenge$). Results for $9_1$ are omitted to improve the clarity of the plot but are seen to fit with the general trend.}
\end{center}
\end{figure}

\begin{figure}
\begin{center}
\includegraphics[scale=0.2]{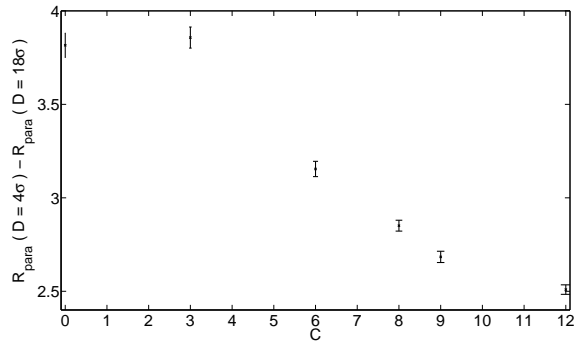}
\caption{\label{fig:R_para_change} Change in the radius of gyration in the two free directions, $R_{para}$, between slits of width $D = 18\sigma$ and $D = 4\sigma$ for different knot types. Knot type is indicated on the horizontal axis by $C$, the essential crossings.}
\end{center}
\end{figure}

Although it is intuitively reasonable that for wider slits more complex knots lead to a more compact polymer and hence a lower monomer density at the walls, it is less obvious why the monomer density at the walls should increase with knot complexity in narrower slits. A suggestion as to why this may be the case is found by looking at the distribution of (instantaneous) $R_{para}$, the radius of gyration in the two unconfined directions, for a slit with $D = 18\sigma$, the widest we consider. As shown in Fig.~\ref{fig:R_para_dists}, for more complex knots the distribution is narrower and has a maximum at lower $R_{para}$ indicating that the more complex knots prevent the polymer from exploring configurations where they are spread out along the slit. Hence this suggests that when the slit becomes narrow and significantly distorts the polymer, the polymer cannot so easily distribute its monomers further out in the free directions and instead their density near the wall increases. As shown in Fig.~\ref{fig:R_para_change} the increase in the average $R_{para}$ as $D$ is changed from $18\sigma$ to $4\sigma$ is less for more complex knots.

\subsection{Summary and Discussion}

To summarize, we have investigated the effect of knot type on the properties of a ring polymer confined in a slit of width $D$. We found that, for wide slits, more complex knots exert lower forces than a linear polymer on the walls. For more narrow slits the opposite occurs: knotting leads to larger forces. The forces are seen to correlate with the monomer densities near the walls. The crossover occurs at different slit widths for different knots. For larger slit width, when the polymers are not significantly deformed, those with more complex knots have a smaller radius of gyration~\cite{quake}, reducing the monomer density by the walls and correspondingly the force exerted on them. For smaller slits, the knot prevents the polymer from deforming as much in the free directions as an unknotted polymer can, and so the monomer densities, and consequently forces, are increased.

The example of $8_{18}$, which has an unusual spherical shape and shows non-monotonic behaviour of the force ratio (to that for a linear polymer) as a function of $D$, demonstrates that the particular topology of the knot plays a role. We conjecture that the maximum for $8_{18}$ occurs because, due to their shape, polymers with $8_{18}$ cannot respond to decreasing $D$ by losing rotational degrees of freedom for their overall configuration, as more elongated polymers can. Instead, they begin to deform at a point, relative to their size, where other polymers still maintain their preferred shape, if with reduced orientational freedom.

Although we exclusively consider one polymer length, the trends we have uncovered should be generic: except in the large $N$ limit, knots make polymers more compact and will decrease their ability to spread out as they are confined. Therefore we expect that there is a wide range of polymer lengths for which similar qualitative trends will be observed. An interesting issue to be explored in the future is the crossover from weak to strong knot localization as the polymer is confined: polymers in a good solvent in three-dimensions are weakly localized~\cite{orlandini2} whilst those in two-dimensions are strongly localized. It would also be useful to look at the crossover from the type of behaviour we observe to the large $N$ limit, where, due to localization, knot type will not affect polymer properties.

\section*{Acknowledgements}

We are very pleased to contribute to this Special Issue of Molecular Physics dedicated to Professor Bob Evans in recognition of his important contributions to the field. We wish him much success in the coming years.

\bibliography{knot_slit_paper}
\bibliographystyle{tMPH} 

\label{lastpage}

\end{document}